# Modeling Smart Grid using Generalized Stochastic Petri Net

[1]Amrita Dey, [2]Nabendu Chaki, [3]Sugata Sanyal

[1, First Author]Techno India College of Technology, India, amritadey84@gmail.com
[*2,Corresponding Author]University of Calcutta, India, nabendu@ieee.org
[3]Tata Institute of Fundamental Research, India, sanyal@tifr.res.in

## Abstract

*Building smart grid for power system is a major challenge for safe, automated and energy efficient usage of electricity. The full implementation of the smart grid will evolve over time. However, before a new set of infrastructures are invested to build the smart grid, proper modeling and analysis is needed to avoid wastage of resources. Modeling also helps to identify and prioritize appropriate systems parameters. In this paper, an all comprehensive model of smart grid have been proposed using Generalized Stochastic Petri Nets (GSPN). The model is used to analyze the constraints and deliverables of the smart power grid of future.*

## 1. Introduction

The electrical power generation, transmission, distribution, and consumption network is referred in this paper as the grid. The function of a grid is an aggregate of multiple networks and multiple power generation companies, employing varying levels of communication and coordination. The risks associated with relying on an overly loaded power grid, exposes us to multiple complex threats. Smart grid is an umbrella term covering modernization of the transmission and distribution grids. Intelligent monitoring system in the smart grid keeps track of all the electricity flows in the system. A smart grid must be able to i) heal itself; ii) motivate consumers to actively participate in grid operations; iii) resist attack; iv) provide higher quality power that saves money; v) offer generation and storage options; and vi) run more efficiently. The full implementation of smart grid will evolve over time. Proper modeling and analysis of smart grid is not only needed for avoiding wastage of resources, but also to identify and isolate security threats.

## 2. Scope of the Work

Some of the recently published works in this field have been studied to develop the Generalized Stochastic Petri Nets (GSPN) model of the Smart Grid. The complex system architecture of smart grid introduces an undetermined level of risk in power systems. Many current techniques to evaluate cyber specific risk fail to scale to such large environments. A model based on access graphs for evaluating the security exposure of a large scale smart grid has been introduced in [1]. Several standardized wired and wireless communication technologies are available for various smart grid applications [2]. The work in [3] further establishes the importance of the smart grid to the current century and discusses a strategy for defining the goals, use cases and system tests so that the complex system meets the needs of the stakeholders. Advanced Metering Infrastructure (AMI) is a very important part of the smart grid [4]. Advanced meters for electricity is focused in [5] for providing two-way communication to upload commands and to download measuring data from the meters. The work in [6] aims at identifying several specific areas of application of computational intelligence in a smart electric grid. In [19], an analytical model using Petri Net has been proposed for distributed data management in a data warehouse to ease the OLAP (Online Analytical Processing) operations.

As evident from the discussion above, very limited attempts have yet been made to model the smart grid for analyzing its constraints and deliverables. No single model has so far been proposed to represent the dynamic and complex behavior associated with smart grid covering all its functionalities spread across generation of power to intelligent billing mechanism. In this paper, an all comprehensive model of smart grid has been proposed using Generalized Stochastic Petri Nets (GSPN). The model has been used to analyze the constraints and deliverables of the smart power grid of future.

## 3. Petri Net and its Variants

Petri nets and its variants, often referred as high level nets, have been used to model and analyze wide range of application area from unmanned aerial vehicle to processor architecture. Multistage Interconnection Network (ICN) has been modeled using another variant of high-level net called S-net [9]. The elementary place time Petri Net (PN) in [12], has been used to analyze performance of the flexible production system. In [10], performance of system is compared on GSPN models by analyzing attack trees. Another extension to the elementary place-transition net is the Colored PN (CPN). A CPN model of an urban traffic network for performance evaluation is presented in [18]. In [11], the interval timed CPN has been defined as a 5-tuple ($\Sigma$, P, T, C, F) such that $\Sigma$ is a finite set of types, also referred to as colors. Multiple colors can be associated with a given type, P is a finite set of places, and T is a finite set of transitions such that P $\cap$ T = $\emptyset$, C is a color function and F is the transition function. PN based software architecture for Unmanned Aerial Vehicle (UAV) simulation has been proposed in [14] where, a CPN based hierarchical architecture for multiple UAVs facilitates rapid prototyping via visual modeling and analysis. In one of the recent works, the Stochastic Petri Nets (SPN) has been used for the analysis of signal transduction pathways for angiogenesis processes [7]. In [17] a PN based pilot behavior model is proposed as part of the knowledge based cockpit assistant system. Petri net based analysis also find connotation in e-commerce workflow access control model [15], and for process discovery that can reproduce the logs under consideration in process mining [8]. Weighted fuzzy petri net has been used in [20] for fault analysis of a Flight Control System.

GSPN provides a graphical representation of a system that can be used for the formal description and stochastic and time dependent analysis of systems [16]. The model of a smart grid should cover the characteristics of a smart grid including self-healing mechanism, accommodating different generation and storage options, supporting the concept of on-site power generation and utilization at the consumer end and use advanced tools and metering infrastructure, etc. in the grid. Besides, the designed model of the smart grid is expected to be bounded, safe and deadlock free for efficient analysis of the system. Keeping all these in mind, GSPN has been selected to model and analyze the smart grid.

## 4. Modeling Smart Grid using GSPN

The GSPN model of the smart grid should cover the characteristics of a smart grid including self-healing mechanism, accommodating different generation and storage options, supporting the concept of on-site power generation and utilization at the consumer end and use advanced tools and metering infrastructure, etc. in the grid. Besides, the designed model of the smart grid is expected to be bounded, safe and deadlock free for efficient analysis of the system.

The smart grid can be thought of as a synchronous series of events starting with smart power generation, transmission, distribution and finally to the smart consumption of the power. The designed smart grid model would be a GSPN having both timed and immediate transitions. The transition delays of the timed transitions are assumed to be random variables with negative exponential distributions. Based on the elementary model of the smart grid, each of the stages from power generation to consumption is expanded to include all the sub-events occurring in that phase.

In this paper, the proposed GSPN model of the smart grid has been simulated using a tool called PIPE2 (Platform Independent Petri Net) [13]. PIPE2, however, is not robust enough to handle a huge number of timed transitions. As the number of timed transitions increases, the number of tangible states also increases. Thus the proposed model must make sure that only a finite number of tangible states are formed for GSPN analysis. At the same time, the inherently complex functionalities of a smart grid mentioned at the beginning of this section should not be left out in the proposed model. A trade off has been done here to keep a balance. As the smart grid is an extremely large electric power infrastructure, it is natural that the model representing the smart grid is pretty complex. Hence, to better understand the model, each phase of the smart grid will be taken up separately. Smart power generation is the first phase in smart grid or as a matter of fact, in any electric grid.

In figure 1, $T_0$, $T_1$, $T_2$, $T_3$ and $T_4$ represent the harnessing of renewable energy from various sources which are infinite. Thus, these five transitions are represented as hanging transitions without any preset. Moreover, these are timed transitions as they harness the energy only when the energy sources are very intense and not whenever the energy source is available. In any electric grid, energy is mainly harnessed from non-renewable energy sources like coal, oil and natural gas. These have been represented by $P_6$, $P_7$ and $P_8$ respectively with a finite number of tokens. The processing of these non-renewable sources of energy, represented by transitions $T_5$, $T_6$ and $T_7$ are timed transitions. These are enabled with the availability of coal, oil and natural gas respectively and are fired only when the amount of energy is above a certain minimum.

The generation of solar, wind, biomass, geothermal and hydropower energy has been shown by $T_8$, $T_9$, $T_{10}$, $T_{11}$ and $T_{12}$ respectively and $T_{13}$ represents the generation of usable energy from them. $T_8$ to $T_{13}$ are all timed transitions. They are all enabled with the availability of energy but are fired only when the amount of energy in the storage ($P_{11}$) falls below a certain minimum. $T_8$ to $T_{13}$ thus have stochastic timings associated with them. The generated energy from all the sources is stored in the centralized energy storage ($P_{11}$). However, as storage can be expensive, energy is generated as per requirement.

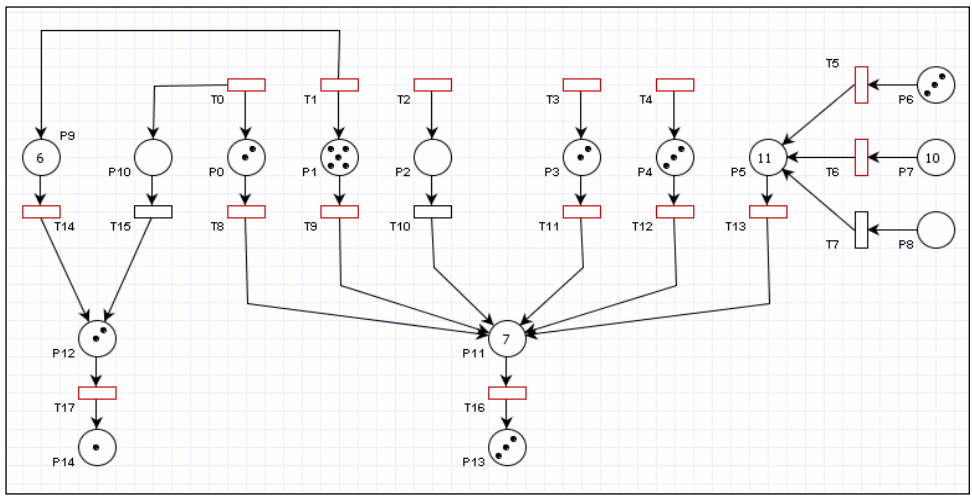

**Figure 1**: GSPN model of smart power generation

The energy at the storage needs to be converted to electricity. This is done by the centralized electricity generator ($P_{13}$). The generation of electricity, shown by $T_{16}$ is again a timed transition. $T_{16}$ is enabled whenever energy is present in the storage but the generation of electricity is based on the demand in the grid. This allows the demand response management as expected from a smart grid.

Smart grid also supports the concept of distributed generation besides renewable energy sources. Distributed generation too has been incorporated in the designed model. Roof-top solar panels ($P_{10}$) and on-site wind turbines ($P_9$) harness the solar and wind energy ($T_0$ and $T_1$) under the same criteria as for the centralized system. Other sources of energy cannot be harnessed on-site as they require huge setup. The energy harnessed onsite is used to generate usable energy ($T_{14}$ and $T_{15}$). Again, these onsite energy generating events are timed as they are fired only when the amount of energy in the distributed energy storage ($P_{12}$) goes below a certain minimum. The stored energy in $P_{12}$ is used to generate on-site electricity using the distributed power generator ($P_{14}$). The electricity generated onsite has to be used as soon as it is generated. This marks the end of the generation phase.

The next phase in line is the smart transmission of the generated power. This phase has been elaborated in figure 2. The generated electricity at the centralized power generator ($P_{13}$) is put to the transmission network. The transmission network consists of multiple transmission lines of superconducting wires to minimize the loss of electricity during transmission and also to increase the transmission capacity. To keep the model simple, only three transmission lines have been shown. Timed transitions $T_{18}$, $T_{19}$ and $T_{20}$ put the generated electricity on to transmission lines 1 ($P_{15}$), 2 ($P_{16}$) and 3 ($P_{17}$) respectively. The generated electricity is transmitted through the line having the maximum requirement. The electricity in the line is transmitted to its utility. At the utility level, advanced sensors are randomly dropped on the transmission line.

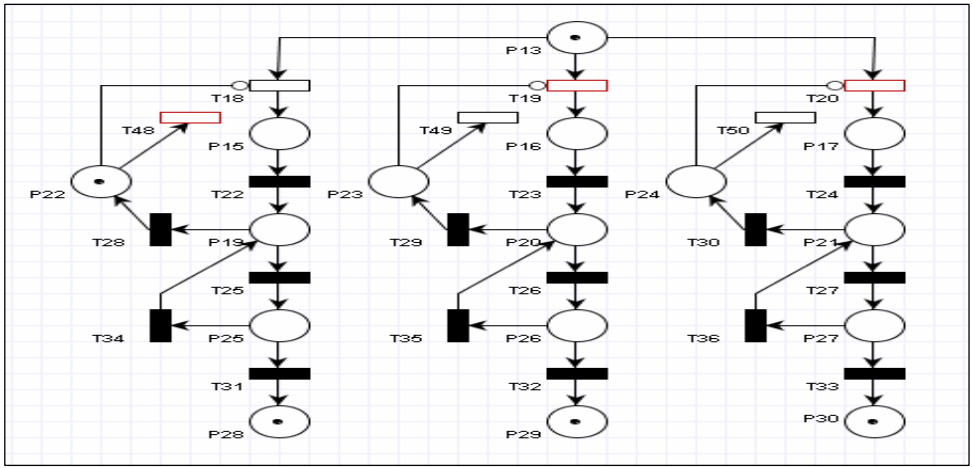

**Figure 2**: GSPN model of smart power transmission

The sensors for the transmission lines 1, 2 and 3 are $P_{19}$, $P_{20}$ and $P_{21}$ respectively. These sensors monitor and report to the visualization tools about real time line conditions. In case the sensors detect any fault, the unit under consideration is removed from the line ($T_{28}$, $T_{29}$ or $T_{30}$). This removed unit of electricity is gathered at a place ($P_{22}$, $P_{23}$ or $P_{24}$) in order to block the defective line temporarily. Inhibitor arcs from $P_{22}$, $P_{23}$ and $P_{24}$ to $T_{18}$, $T_{19}$ and $T_{20}$ have been used for the purpose. $T_{18}$, $T_{19}$ and $T_{20}$ can fire only if there is no token at places $P_{22}$, $P_{23}$ and $P_{24}$. Thus, if the sensors detect overload or fault in its line, then that line cannot be used for transmitting electricity till the fault is repaired or the load is balanced. The line is cleared by the firing of transition $T_{48}$, $T_{49}$ or $T_{50}$. Had the unit of electricity been faulty and put to place $P_{22}$, $P_{23}$ or $P_{24}$ after having being removed from the grid, then $T_{48}$, $T_{49}$ or $T_{50}$ is fired immediately so that the transmission line is not blocked. $T_{48}$, $T_{49}$ and $T_{50}$ are timed transitions as due to faulty line conditions, it is possible that these transitions are enabled but are not fired till the lines are cleared. The sensors report real time line conditions to the visualization tools or remove the transmitting unit of electricity from the grid on the face of any fault. As the sensors are always active, these are immediate transitions. The visualization tools and technologies provide wide area grid awareness and rapid information on black out and power quality of the transmission line. These tools are always active and hence $T_{31}$, $T_{32}$ and $T_{33}$ are immediate transitions.

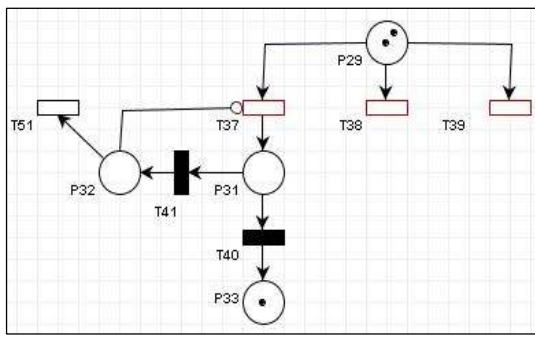 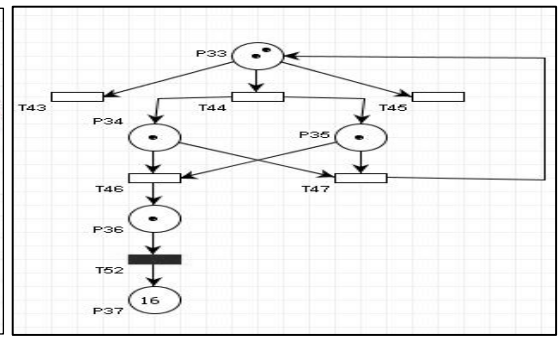

**Figure 3:** Model for smart power distribution  **Figure 4**: Model for smart power consumption

The distribution of electricity has been shown in the figure 3. Each transmission line breaks up into multiple distribution lines before the electricity reaches the end users. Again, to keep the model simple, only one of the multiple transmission lines (line 2), has been broken into only three distribution lines. From the utility ($P_{29}$), the transmission line 2 is shown to break up into three distribution lines. The event of distributing the electricity to the three distribution lines have been shown by the three timed transitions $T_{37}$, $T_{38}$ and $T_{39}$. These transitions are enabled as soon as at least one unit of electricity reaches the utility. However, which of the enabled transition would fire and at what time depends on several factors including the requirements of the customers. Once the distribution line is chosen and before the electricity reaches the substation, it has to go through the Phasor Measurement Unit ($P_{31}$). The PMU measures the power quality ($T_{40}$) in order to control congestion. The electricity then reaches the substation ($P_{33}$). The next phase of smart power consumption, has been shown in Fig 4. A smart grid customer is represented by a set of smart appliances ($P_{34}$) and/or plug-in electric vehicles ($P_{35}$). The pattern of usage of electricity is monitored ($T_{52}$) using an Advanced Metering Infrastructure ($P_{36}$). $T_{52}$ is an immediate transition.

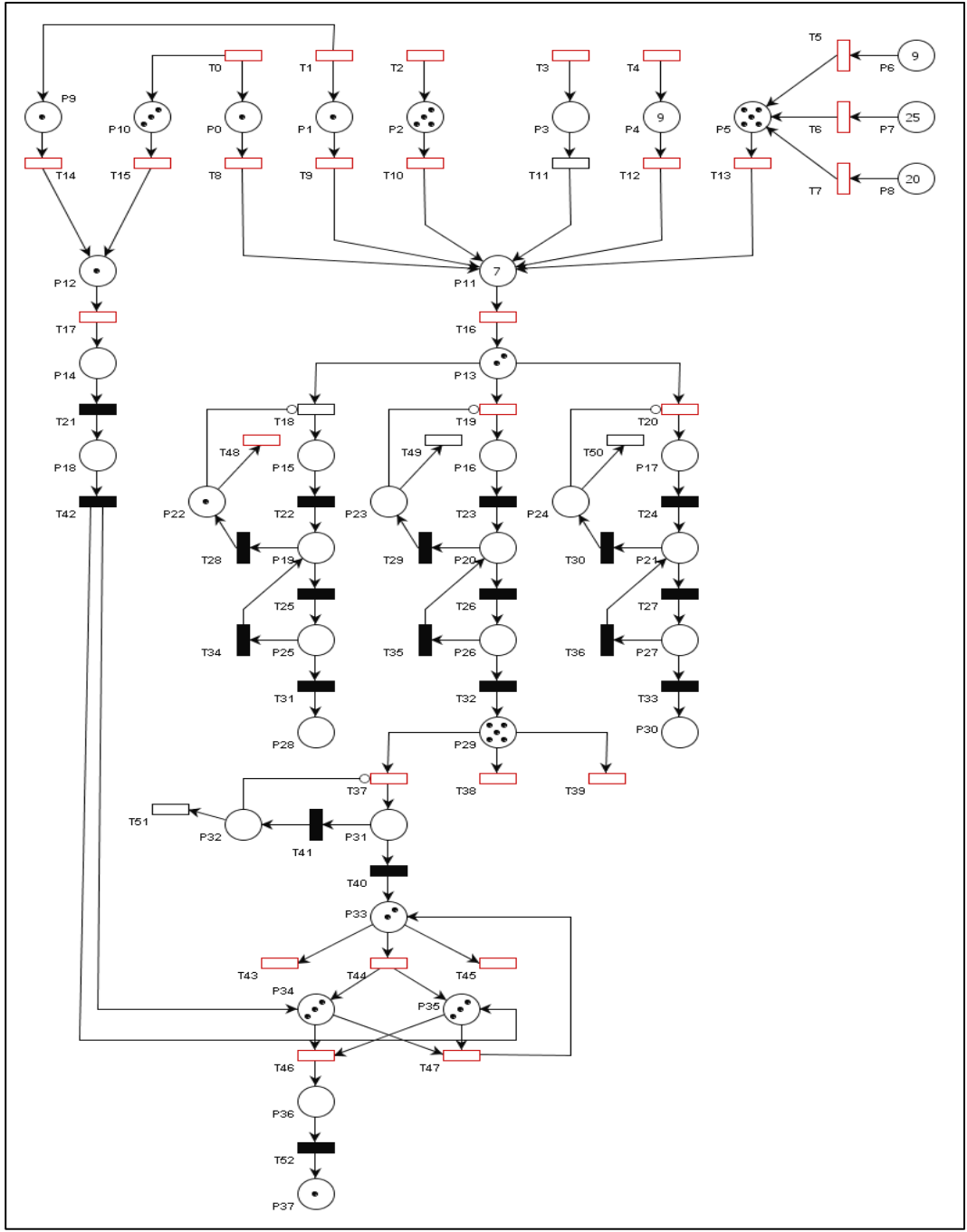

**Figure 5**: Screen-shot of the GSPN Model for a complete smart grid

The AMI not only records the amount of electricity used but also the time of use of the electricity so that this information can be used to charge a premium to those customers using electricity during peak hours. The bill of the customer ($P_{37}$) is thus prepared using this information. The more the number of tokens ($P_{37}$), higher is the number of chargeable units used by the customer. The electricity that has been generated at the customer's site using the distributed power generator ($P_{14}$), is put to the grid ($T_{21}$) as soon as it is generated, thus $T_{21}$ is an immediate transition. The electricity is then passed through the distributed energy resources (DER) based micro-grid ($P_{18}$) that reduces the load ($T_{42}$) before transferring it to the smart consumer appliances, etc.

The screen shot of the GSPN model for the entire smart grid simulated in PIPE2 is shown in figure 5. It is found that all the features and technologies of the smart grid have been successfully incorporated in the model. One important point worth mentioning here is that, the explanations provided for the selection of a timed transition for firing from the set of multiple enabled transitions, are absolutely logical. In collaboration with some electric company, each timed transition may be provided with a fixed rate. The rates would then be used to generate the probability of firing of each timed transition.

## 5. Simulation of the designed GSPN model of Smart Grid in PIPE2

The designed GSPN model of the smart grid has been simulated in PIPE2 using different numbers of initial random firings. The distribution of tokens has been recorded for 100, 300, 500, 700 and 1000 random firings. Again for each of these numbers, the model has been simulated five times to note the variation of the token distribution with the same number of initial random firings. The averages of the five sets (i.e., the average of Series 1, 2, 3, 4 and 5) for each of the five values of initial random firings (i.e., for 100, 300, 500, 700, 1000) have been used for the final analysis. The broken simulations of the model, with different number of initial random firings result in different distribution of tokens in the model. The simulation results have been presented graphically to aid the analysis. The token distribution has been studied to verify the similarity of the model to the actual smart grid. The averages for the five sets of simulation are shown in Fig 6 to compare the effect on token distribution.

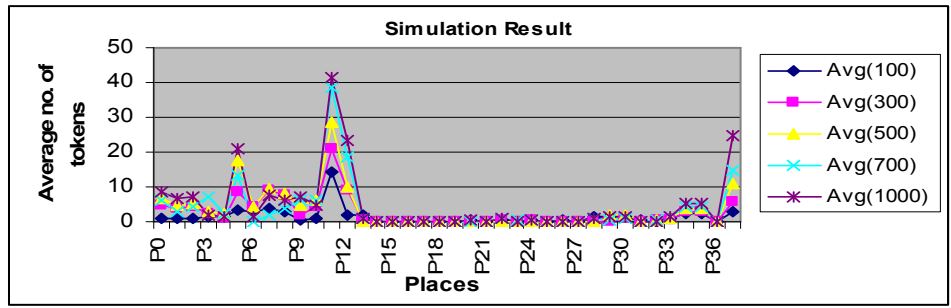

**Figure 6**: Token distribution for 100, 300, 500, 700 and 1000 initial random firings

The graph clearly shows that each line reaches the peak at $P_{11}$ as this place represents the centralized energy storage. This peak is significantly higher than all other peaks in the graph as $P_{11}$ stores the energy harnessed and processed from the different renewable and non-renewable sources. The distributed energy storage ($P_{12}$) also shows a high for each of the five lines, though the high is much less than the centralized energy storage, since, under normal conditions, the energy harnessed onsite is much less compared to that harnessed for the centralized power generator. Energy is stored at both the centralized and the distributed energy storage till there is demand from the customer end for electricity. One unit of energy is taken at a time from these storages and electricity is generated at the centralized ($P_{13}$) or the distributed power generator ($P_{14}$).

The maximum token density for all the five lines in the above graph is observed at places $P_0$ to $P_{12}$ that forms the generation phase. The varying token densities at the PV panels ($P_0$ and $P_{10}$), wind turbines ($P_1$ and $P_9$), biomass burner ($P_2$), geo-thermal well ($P_3$) and hydropower plant ($P_4$) are absolutely random and depend on the intensity of the various non-perishable sources of energy. Moreover, the tokens from these places are consumed to generate energy only when the units of energy stored at the centralized and distributed energy storage fall below a certain minimum. This is because, though energy can be stored in the smart grid, this could be expensive. All five lines are significantly high at $P_5$, the storage for processed coal, oil and natural gases. Relatively lower values are found at $P_6$, $P_7$ and $P_8$ that represent the storages for raw coal, oil and natural gases. This again proves that though the smart grid supports renewable energy sources, the bulk of energy is harnessed from the traditional perishable sources. The token density at the transmission and distribution lines appears to be almost zero. This is because, once the electricity is generated, there is no scope of storing the electricity.

The token density rises at $P_{34}$ and $P_{35}$ which represent the units of electricity being used by the customers. In case the received electricity is higher than that demanded by the customer, the surplus electricity is sold back to the substation ($P_{33}$). Hence, multiple units of electricity can get accumulated at the substation at a particular instance. Thus, a very small peak is seen at $P_{33}$ (substation), if observed carefully. While the electricity is being used by the customer, the advanced metering infrastructure ($P_{36}$) records the usage pattern to prepare the electricity bill of the customer. The high token density at $P_{37}$ (bill) represents the number of chargeable units of electricity used by that customer.

In general, it is clear from figure 6 that higher the number of initial random firings, higher is the peaks reached. The graphs clearly show that the line representing 1000 initial random firings has the highest token density in all the four phases as compared to the other series. In order to offer better clarity of the explanation, the graph in Fig 7 has been broken into four sub-graphs according to the four phases of the smart grid. Figs 8 to Fig 11 show the graphs specific to each of the phases.

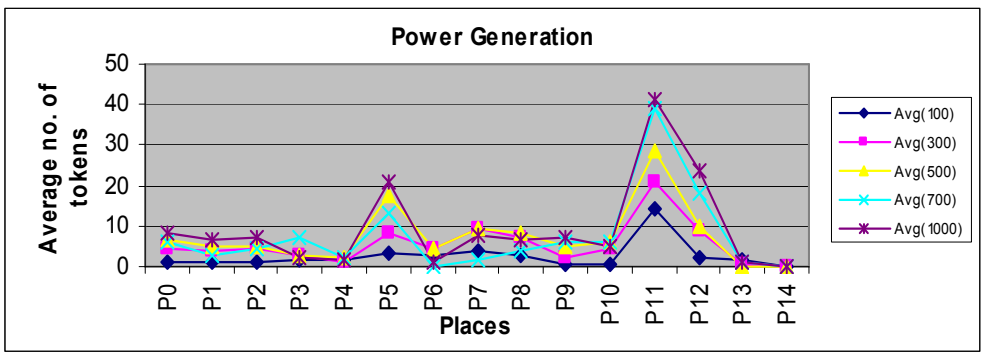

**Figure 7**: Average token distribution in the generation phase different initial random firings

In Fig 7, there are two distinct peaks. The highest and the second highest peaks for all the five different initial firings are of course at the centralized energy storage ($P_{11}$) and the storage for processed coal, oil and natural gases ($P_5$). Places $P_6$ (raw coal), $P_7$ (raw oil) and $P_8$ (raw natural gases) have higher token density then places $P_0$ to $P_4$, indicating that quantity of perishable energy harnessed is much more than the quantity of renewable energy harnessed.

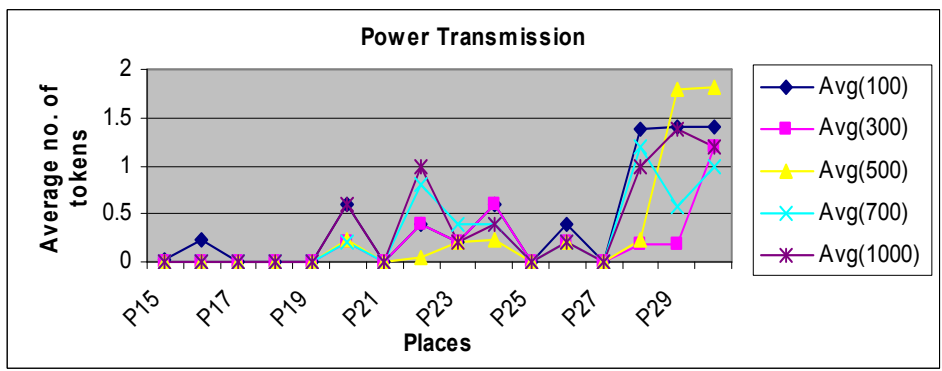

**Figure 8**: Average token distribution in transmission phase for different initial random firings

The distributed energy storage ($P_{12}$) also has a high density, though lesser than that at the centralized energy storage. The densities of tokens at $P_{15}$ to $P_{27}$ are 1 or close to 1. High peaks are observed at $P_{28}$,

$P_{29}$ and $P_{30}$ (utilities). However, a careful study of Fig 7 and Fig 8 shows that while the highest peak in Fig 7 is more than 40, the highest peak in Fig 8 is less than 2. The reason for such low token density of all the places in the transmission phase is that, the generated electricity cannot be stored.

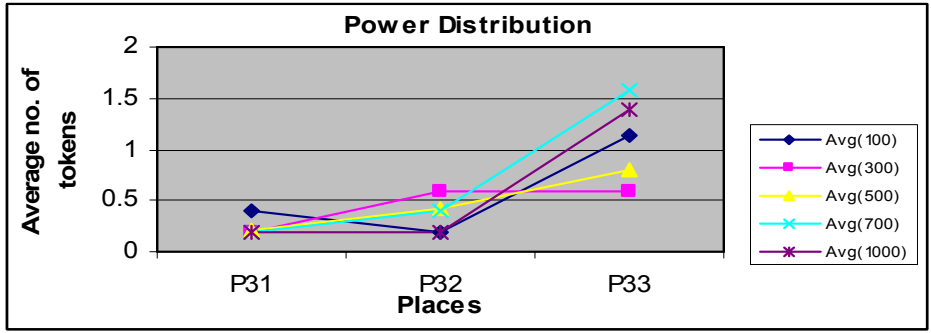

**Figure 9**: Average token distribution in distribution phase for different initial random firings

In fig 9, the token densities at $P_{31}$ and $P_{32}$ are consistently below 0.6. However, the token density at place $P_{33}$ (substation) rises a little, though nowhere close to the high obtained at the centralized energy storage in Fig 7. At a given instance, the substation can have more than one unit of electricity as it not only gets electricity from the distribution lines but also buys electricity from the customers.

The behavior of the graph in the consumption phase, as shown in Fig 10, is pretty obvious. The token densities at $P_{34}$ and $P_{35}$ represent the units of electricity that the customer had been using at the instance the simulation had been done. As the AMI ($P_{36}$) records the usage pattern of one unit of electricity at a time, this place has low token density. The average token densities of this place, for the five different values of initial random firings, are between 0 and 1. The high at the electric bill ($P_{37}$) represents the chargeable units of electricity used by the customer throughout the month.

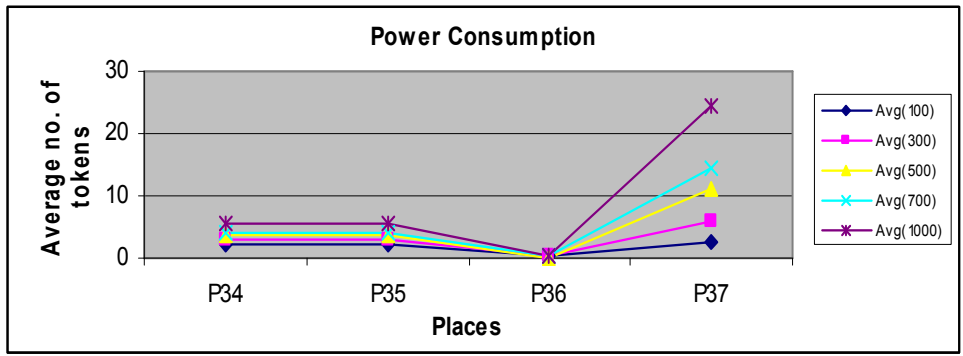

**Figure 10**: Average token distribution in consumption phase for different initial random firings

The study of the graphs shows that, while the generation phase has the maximum token density, the consumption phase shows the second highest token density. This is because the tokens represent units of energy or electricity, as applicable. Now, once electricity is generated, it cannot be stored. Thus each place from the centralized power generator to the substation cannot have more than one token at a time. The junctions between the various phases, i.e., the place marking the end of one phase and the beginning of another phase, can have more than one token at any particular instance. Few tokens can get accumulated at these places due to the unequal rates of electricity generation, transmission, distribution and consumption. Large token density at the final place of the consumption phase, the electricity bill ($P_{37}$) indicates total chargeable units used by customer throughout the month.

The model clearly shows that transmission and distribution cannot begin till electricity has been generated. Again, electricity cannot be generated unless some renewable or nonrenewable source of energy is available. However, if the transmission or distribution network fails for some reason, then, the consumers are able to generate sufficient electricity onsite using Smart grid technologies like DG and DER. Another point worth noting is that the AMI helps the consumers to better use the received electricity to reduce their electricity bills. Had the AMI not been there in the system, the smart grid would have successfully delivered the generated power to the consumers but it would have behaved more or less like our current grid. The addition of the advanced sensors and the visualization tools at the utility level allows the smart grid to monitor the transmission to provide rapid information about power outages and power quality. Phasor measurement unit provides a dynamic view of the system.

## 6. Future Scope and Conclusion

Though the smart grid will be the electric power grid of our future, no attempt has yet been made to model the smart grid. The GSPN model of the smart grid, designed in this paper, has laid the foundation for the functional and performance analysis of the smart grid. The simulation of the model has allowed the analysis of how close the behavior of the model is to the behavior of the actual smart grid. The average distribution of tokens has been studied and compared for different number of initial random firings. Since, the designed model has proved to support all the features of the smart grid and incorporate all the technologies used in the smart grid; the model can be used by the smart grid engineers and designers for what-if analysis and all kinds of experiments. The model could have been broken according to its four phases and the stochastic analysis of the phases could have been performed separately. However, for proper GSPN analysis, valid rates need to be assigned to the timed transitions and accurate weights to immediate transitions. This could have been done only if collaboration with some electric company had been possible. In future, the stochastic analysis of the designed model may be done by working in collaboration with some electricity company like Calcutta Electric Supply Corporation (CESC).

## 7. References

[1] A. Hahn, M. Govindarasu; Smart Grid Cyber Security Exposure Analysis and Evaluation Framework; Power and Energy Society General Meeting, IEEE; ISSN: 978-1-4244-6551-4/10; 2010.
[2] P. P. Parikh, M. G. Kanabar, T. S. Sidhu; Opportunities and Challenges of Wireless Communication Technologies for Smart Grid Applications; IEEE Power and Energy Society General Meeting; ISSN: 978-1-4244-6551-4/10; 2010.
[3] E. Simmon, A. Griesser; Use Case Based Testing to Improve Smart Grid Development; New World Situation: New Directions in Concurrent Engineering, DOI: 10.1007/978-0-85729-024-3_57; 2010.
[4] L. Zhang; Study of Applications Based on Measurement Technology in the Future Smart Grid; Communications in Computer and Information Science; 1; Vol. 93; Advanced Intelligent Computing Theories and Applications; Part 24; pp. 493-498; 2010.
[5] G. Deconinck; Metering, Intelligent enough for Smart Grids?; Topics in Safety, Risk, Reliability and Quality, 1, Volume 15, Securing Electricity Supply in the Cyber Age, Pages 143-157; 2010.
[6] Z. Jiang; Computational Intelligence Techniques for a Smart Electric Grid of the Future; Lecture Notes in Computer Science, Vol. 5551, Advances in Neural Networks – ISNN; pp 1191 – 1201; 2009.
[7] L. Napione et. al.; On the Use of Stochastic Petri Nets in the Analysis of Signal Transduction Pathways for Angiogenesis Process; Int'l conf. on Computational Methods; pp. 281-295; 2009.
[8] N. Busi, G. M. Pinna, Process discovery and Petri nets; Journal of Mathematics Structure in Computer Science, Cambridge Univ. Press, vol. 19, pp. 1091–1124; 2009.
[9] N. Chaki, S. Bhattacharya, Performance analysis of multistage interconnection networks with a new high level net model, Journal of System Architecture; Elsevier North-Holland, Inc., vol. 52(1), pp. 56–70, ISSN: 1383-7621, 2006.
[10] G. C. Dalton II, J. M. Colombi; Analyzing Attack Trees using Generalized Stochastic Petri Nets, Proc. of IEEE workshop on Information Assurance, pp.116–123, ISBN: 14244-01305, 2006.
[11] O. M. Dahl, S. D. Wolthusen; Modeling and Execution of Complex Attack Scenarios using Interval Timed Colored Petri Nets; 4th IEEE Int'l workshop on Info. Assurance; pp. 157–168, 2006.
[12] A. W. Colombo, R. Carelli, B. Kuchen; A Temporized Petri Net Approach for Design, Modeling and Analysis of Flexible Production Systems, International Journal of Advanced Manufacturing Technology, Springer, vol. 13(3); pp. 214–226, ISSN: 0268-3768, 2005.
[13] Nadeem Akharware: PIPE2: Platform Independent Petri Net Editor, University of London, 2005

[14] D. Xu, et. al., A Petri Net Based Software Architecture for UAV Simulation, Proc. of the Int'l conf. on Software Engineering Research and Practice, pp. 227-234, 2004.
[15] C. Zhuo, et. al.; Petri Net Based Workflow Access Control Model; Journal of Shanghai Univ; pp. 63-69, vol. 8(1), 2004.
[16] M. Ajmone, et. al.; Modeling with Generalized Stochastic Petri Nets; Wiley Series in Parallel Computing, 1998.
[17] W. Ruckdeschel, R Onken: Modeling of pilot behavior using Petri Nets, 15th Int'l conf. on Application and Theory of Petri Nets, Spain, pp. 436-453, 1994.
[18] F. Dicesare, et. al.: The Application of Petri Nets to the Modeling, Analysis and Control of Intelligent Urban Traffic Networks; Int'l conf. on Application and Theory of Petri Nets, pp. 2-15, 1994.
[19] Bidyut Biman Sarkar, Nabendu Chaki; "Virtual Data Warehouse Modeling Using Petri Nets for Distributed Decision Making"; JCIT, Vol. 5(5), pp. 8-21, July 2010. ISSN: 1975-9320, 2010.
[20] Jiufu LIU, Kui Chen, Zhisheng WANG, "Fault Analysis for Flight Control System Using Weighted Fuzzy Petri Nets", JCIT, Vol. 6, No. 3, pp. 146-155, 2011.

**Acknowledgement:** *The authors thankfully acknowledge the Department of Computer Science & Engineering, University of Calcutta, India for the institutional support in carrying out this work. The code for simulation may be requested at nchaki@cucse.org for academic use and citation subject to the prior consent of the corresponding author.*

AUTHOR BIOGRAPHY

**Amrita Dey** is an Assistant Professor in the Computer Science & Engineering Department of Techno India College of Technology, Rajarhat, Kolkata, affiliated to West Bengal University of Technology. She has done B.Sc. in Computer Science from St. Xavier's College, Kolkata and her M. Sc and M. Tech. in Computer Science are from the University of Calcutta. Amrita has been a part-time Lecturer of Computer Science in Lady Brabourne College, Kolkata. Her research interest includes distributed systems. Amrita has particular interest in simulation software and has handled several simulation platforms.

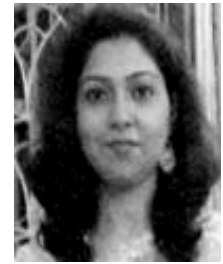

Nabendu Chaki is HoD and an Associate Professor in the Department Computer Science & Engineering, University of Calcutta, Kolkata, India. He did his double graduation in Physics and in Computer Science & Engineering, both from the University of Calcutta. Dr. Chaki has completed Ph.D. in 2000 from Jadavpur University, India. He has authored a couple of text books and close to 90 refereed research papers in Journals and International conferences. His areas of research interests include distributed computing and software engineering. Dr. Chaki has also served as a Research Assistant Professor in the Ph.D. program in Software Engineering in U.S. Naval Postgraduate School, Monterey, CA. He is a visiting faculty member for many Universities including the University of Ca'Foscari, Venice, Italy. Dr. Chaki is a Knowledge Area Editor in Mathematical Foundation for the SWEBOK project of the IEEE Computer Society. Besides being in the editorial board for five International Journals, he has also served in the committees of close to 50 international conferences.

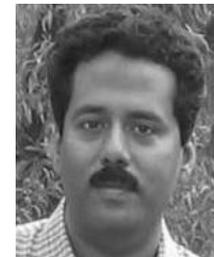

Sugata Sanyal is a Professor in the School of Technology & Computer Science at the Tata Institute of Fundamental Research (http://www.tifr.res.in/~sanyal) and is working there from 1973. After an extensive stint in building Computer and also Computerized Systems during '70s and '80s, he has worked in diverse areas of Computer Architecture, Parallel Processing, Fault Tolerance and Coding Theory and in the area of Security. Details of his work may be found from his homepage. He has been associated with many conferences in India and abroad. His publication List is also available in his homepage. Sanyal is in the Editorial Board of many International Journals, and is collaborating with scientists from India and abroad.

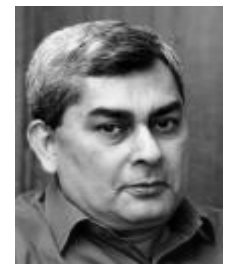